\documentclass[12pt,preprint]{aastex}

\newcommand{\Li}{{\rm Li}}

\shorttitle{Substellar Mass Limit}
\shortauthors{Auddy et al.}

\begin{document}


\title{ Analytic Models of Brown Dwarfs and The Substellar Mass Limit }

\author{Sayantan Auddy\altaffilmark{1}, Shantanu Basu\altaffilmark{1}, and S. R. Valluri\altaffilmark{1,2}}

\altaffiltext{1}{Department of Physics and Astronomy, The University of Western Ontario, London, ON N6A 3K7, Canada.}
\altaffiltext{2}{King's University College, The University of Western Ontario, London, ON N6A 2M3, Canada.}
\email{sauddy3@uwo.ca, basu@uwo.ca, valluri@uwo.ca.}






\begin{abstract}
We present the current status of the analytic theory of brown dwarf evolution and the lower mass limit of the hydrogen burning main sequence stars. In the spirit of a simplified analytic theory we also introduce some modifications to the existing models. We give an exact expression for the pressure of an ideal non-relativistic Fermi gas at a finite temperature, therefore allowing for non-zero values of the degeneracy parameter ($\psi = \frac{kT}{\mu_{F}}$, where $\mu_{F}$ is the Fermi energy). We review the derivation of surface luminosity using an entropy matching condition and the first-order phase transition between the molecular hydrogen in the outer envelope and the partially-ionized hydrogen in  the inner region. We also discuss the results of modern simulations of the plasma phase transition, which illustrate the uncertainties in determining its critical temperature. Based on the existing models and with some simple modification we find the maximum mass for a brown dwarf to be in the range $0.064M_\odot-0.087M_\odot$. An analytic formula for the luminosity evolution allows us to estimate the time period of the non-steady state (i.e., non-main sequence) nuclear burning for substellar objects. Standard models also predict that stars that are just above the substellar mass limit can reach an extremely low luminosity main sequence after at least a few million years of evolution, and sometimes much longer. We estimate that $\simeq 11 \%$ of stars take longer than $10^7$ yr to reach the main-sequence, and $\simeq 5 \%$ of stars take longer than $10^8$ yr.  
\end{abstract}

\newpage

\keywords{equation of state -- stars: fundamental parameters -- stars: interiors -- stars: low-mass, brown dwarfs -- stars: pre-main sequence}



\section{Introduction}
One of the most interesting avenues in the study of stellar models lies in understanding the physics of objects at the bottom of and below the hydrogen burning main sequence stars. The main obstacle in the study of very low mass (VLM) stars and substellar objects is their low luminosity, typically of order $ ~10^{-4} L_\odot$, which makes them difficult to detect. There is also a degeneracy between mass and age for these objects, which have a luminosity that decreases with time. This makes a determination of the initial mass function (IMF) difficult in this mass regime. However in the last two decades there has been substantial observational evidence that supports the existence of faint substellar objects. Since the first discovery of a brown dwarf \citep{Opp95,Reb95} several similar objects were identified in young clusters \citep{Mat96} and Galactic fields \citep{Ruiz97} and have generated great interest among theorists and observational astronomers. The field has matured remarkably in recent years and recent summaries of the observational situation can be found in \citet{Luhman07} and \citet{Chab14}.


In two consecutive papers, \citet{Kum63} and \citet{Kum263} revolutionized the understanding of low mass objects by studying the Kelvin-Helmholtz time scale and structure of very low mass stars. He successfully estimated that stars below about $0.1M_\odot$ contract to a radius of about $0.1R_\odot$ in about $10^9$ years, which was a correction to the earlier estimate of $10^{11}$ years. The earlier calculation was based on the understanding that low mass stars evolve horizontally in the H-R diagram and thus evolve with a low luminosity for a long period of time. However, \citet{Hay78} showed that such low mass stars remain fully convective during the pre-main sequence evolution and are much more luminous than the previously accepted model based on radiative equilibrium. Kumar's analysis showed that for a critical mass of $0.09M_\odot$, the time scale has a maximum value that decreases on either side. Although this crude model neglected any nuclear reactions, it did give a very close estimate of the time scale.
The second paper \citep{Kum263} gave  more detailed insight on the structure of the interior of low mass stars. This model was based on the non-relativistic degeneracy of electrons in the stellar interior. Kumar's extensive numerical analysis for a particular abundance of hydrogen, helium and other chemical compositions yielded a limiting mass below which the central temperature and density are never high enough to fuse hydrogen. A more exact analysis required a detailed understanding of the atmosphere and surface luminosity of such contracting stars.

The next major breakthrough in theoretical understanding came from the work of  \citet{Hay78}, who studied the pre-main sequence evolution of low mass stars in the degenerate limit.
Although it was predicted that there exist low mass objects that cannot fuse hydrogen, the internal structure of these objects remained a mystery.
A complete theory demanded a better understanding of the physical mechanisms which govern the evolution of these objects. It became essential to develop a complete equation of state (EOS). 

\cite{Ant85} used numerical simulations to study the evolution of VLM stars and brown dwarfs for Population I chemical composition ($Y = 0.25,\, Z = 0.02 $) and different opacities.  Their model showed that for the same central condition (nuclear output) an increasing opacity reduces the surface luminosity.  Thus, a lower opacity causes a greater surface luminosity and subsequent cooling of the object. Their results implied that the hydrogen burning minimum mass is $M=0.08 M_{\odot}$ for opacities considered in their model. Furthermore, they showed that objects with mass close to  $M=0.08 M_{\odot}$ spend more than a billion years at a luminosity of $\sim 10^{-5} L_{\odot}$. 

\citet{Burs89} modeled the structure of stars in the mass range $0.03M_\odot-0.2 M_\odot$. They used a detailed numerical model to study the effects of varying opacity, helium fraction, and the mixing length parameter, and compared their results with the existing data. Their important modification was that they considered thermonuclear burning at temperatures and densities relevant for low masses. A detailed analysis of the equation of state was performed in order to study the thermodynamics of the deep interior, which contained a combination of pressure ionized hydrogen, helium nuclei, and degenerate electrons. This analysis clearly expressed the transition from brown dwarfs to very low mass stars. These two families are connected by a steep luminosity jump of two orders of magnitude for masses in the range of $0.07M_\odot-0.09M_\odot$. There was a clear indication that masses in that intermediate regime do ignite hydrogen but that it eventually subsides to yield a brown dwarf. 

\citet{Saumon89} proposed a new EOS for fluid hydrogen that, in particular, connects the low density limit of molecular and atomic hydrogen to the high density fully pressure-ionized plasma. They used the consistent free energy model but with the added prediction of a first order ``plasma phase transition" (PPT) \citep{Saumon89} in the intermediate regime of the molecular and the metallic hydrogen. As an application of this EOS, they modelled the evolution of a hydrogen and helium mixture in the interior of Jupiter, Saturn, and a brown dwarf \citep{Chab& horn92,Chab92}. They adopted a compositional interpolation between the pure hydrogen EOS and a pure Helium EOS to obtain a H/He mixed EOS. This was based on the additive volume rule for an extensive variable \citep{Fon77} and allowed calculations of the H/He EOS for any mixing ratio of hydrogen and helium. Their analysis suggested that the cooling of a brown dwarf with a PPT proceeds much more slowly than in previous models \citep{Burs89}.

\citet{Ste91} presented a detailed theoretical review of brown dwarfs. 
His simplified EOS related pressure and density for degenerate electrons and for ions in the ideal gas approximation. Although corrections due to Coulomb pressure and exchange pressure are  of physical relevance, they together contribute less than $10 \% $ in comparison to the other dominant term in the pressure-density relationship for massive brown dwarfs ($  M \geq 0.04\, M_\odot $). The theoretical analysis gave a very good understanding of the behavior of the central temperature $T_c$ as a function of radius and degeneracy parameter $\psi$. \citet{Ste91} discussed the thermal properties of the interior of brown dwarfs and provided an approximate expression for the entropy in the interior and in the atmosphere of a brown dwarf. He also derived an expression for the effective temperature as a function of mass.

A method to use the surface lithium abundance as a test for brown dwarf candidates was proposed by \cite{Reb92}. Lithium fusion occurs at a temperature of about $2.5 \times 10^{6}\, \rm{K} $, which is easily attainable in the interior of the low mass stars. However brown dwarfs below the mass of $0.065 M_{\odot}$ never develop this core temperature. They will then have the same lithium abundance as the interstellar medium  independent of their age. However, for objects slightly more massive than $ 0.065 M_{\odot}$, the core temperature can eventually reach $3 \times\, 10^{6} \,\rm{K}$. They deplete lithium in the core and the entire lithium content gets exhausted rapidly due to the convection. This causes significant change in the observable photospheric spectra. Thus lithium can act as a brown dwarf diagnostic \citep{Bas96} as well as a good age detector \citep{Stauf96}. 

Following this, an extensive review on the analytic model of brown dwarfs was presented by \citet{Burs93}. They presented an elaborate discussion on the atmosphere and the interior of brown dwarfs and the lower edge of the hydrogen burning main sequence. Based on the convective nature of these low mass objects, they modelled them as polytropes of order $n=1.5$. Once again the atmospheric model was approximated based on a matching entropy condition of the plasma phase transition between molecular hydrogen at low density and ionized hydrogen at high density. The polytropic approximation enabled the calculation of the nuclear burning luminosity within the core adiabatic density profile \citep{Fowler64}. While the luminosity did diminish with time in the substellar limit, the model did show that brown dwarfs can undergo hydrogen burning for a substantial period of time before it eventually ceases. The critical mass deduced from this model did indeed match that obtained from more sophisticated numerical calculations \citep{Burs89}.

In this work we give a general outline of the analytic model of the structure and the evolution of brown dwarfs. We advance some aspects of the  existing analytic model by introducing a modification to the equation of state. We also discuss some of the unresolved problems like estimation of the surface temperature and the existence of PPT in the brown dwarf environment. Our paper is organized as follows. In Section 2 we discuss the derivation of a more accurate equation of state for a partially degenerate Fermi gas. {We incorporates a finite temperature correction to the expression for the Fermi pressure to give  a more general solution to the Fermi integral. In Section 3 we discuss the scaling laws for various thermodynamic quantities for an analytic polytrope model of index $n=1.5$. In Section 4 we discuss the derivation of the equations \citep{Burs93} connecting the photospheric (surface) temperature with density, where the entropies at the interior and the exterior are matched using the  first order phase transition. In the spirit of an analytic model we derive simplified analytic expressions for the specific entropies above and below the PPT.  We also highlight the need to seek alternate methods given current concerns about the relevance of the PPT in BD interiors.  We discuss the nuclear burning rates for low mass objects in Section 5 and determine the nuclear luminosity $L_N$ \citep{Fowler64}. In Section 6 we estimate the range of minimum mass required for stable sustainable nuclear burning. In Section 7, we discuss a cooling model and examine the evolution of photospheric properties over time. In Section 8, we estimate the number fraction of stars that enters the main sequence after more than a million years. In the concluding section, we discuss further possibilities for an improved and generalized theoretical model of brown dwarfs.

\section{Equation of state}\label{2}
In main sequence stars, the thermal pressure due to nuclear burning balances the gravitational pressure and the star can sustain a large radius and nondegenerate interior for a long period of time. However substellar objects like brown dwarfs fail to have a stable hydrogen burning sequence and instead derive their stability from electron degeneracy pressure. A simple but accurate model needs to have a good equation of state that incorporates the degeneracy effect and the ideal gas behaviour at a relative higher temperature. \citet{Burs93} give a pressure law that applies to both the extremes but has a poor connection in the intermediate zone. \citet{Ste91} also gives an empirical relation for the pressure that does include an approximate correction term to connect the two extremes.
Here, in order to obtain a more accurate analytic expression for the pressure, we integrate the Fermi-Dirac integral exactly using the polylogarithm functions $\Li_s (x)$. The most general expression for the pressure is   


\begin{equation}\label{pressure equation}
P_F=g_{s}\int^{\infty}_{0}\frac{4 \pi p^{2}}{(2 \pi\hbar)^3}dp \left(\frac{1}{e^{\beta(\epsilon-\mu)}+b}\right)\left(\frac{1}{3}p\frac{d\epsilon}{dp}\right)
\end{equation}\\
\citep{pad99}, where $b=1$ for the Fermi gas and $\epsilon(p)=\sqrt{p^2c^2 +m^2c^4}-mc^2$ and the other variables are the standard constants. For substellar objects, the electrons are mainly non-relativistic due to relatively low temperature and density. In the non-relativistic limit, i.e. $m^2c^4\gg p^2c^2$, the energy density reduces to $\epsilon(p) \simeq\frac{p^2}{2m} $.
Now rewriting  Eq. (\ref{pressure equation}) in terms of the energy density  
gives
\begin{equation}\label{eqn:pressure equation2}
P_F=a\int_{0}^{\infty} \frac{\epsilon^{\frac{3}{2}}d\epsilon}{e^{\beta(\epsilon-\mu)}+1} ,
\end{equation}
\\
where $ a=\frac{2}{3}\frac{4\pi(2m)^{\frac{3}{2}}}{(2\pi\hbar)^3}$, $\beta=(k_B T)^{-1}$, and we have taken $g_s =2$.
In the limit $T \rightarrow0$, for all $\epsilon < \mu$ the argument of the exponential is negative and hence the exponential goes to zero as $\beta \rightarrow \infty $. Thus the integral reduces to the Fermi pressure at zero temperature. However in a physical situation at finite temperature, the integral can be solved analytically using the polylogs. The details of the exact derivation for a general Fermi integral are shown in  Appendix A. The expression for the pressure of a degenerate Fermi gas at finite temperature is 
\begin{eqnarray}\label{pressure eqn3}
P_F\simeq a\frac{2}{5}\mu^{\frac{5}{2}}-\frac{1}{8}a\beta^{-1}\mu^\frac{3}{2}\ln(1+e^{-\beta\mu})+\\ \nonumber \frac{3}{2}a\beta^{-2}\mu^\frac{1}{2} \frac{\pi^2}{6}+\frac{3}{4}a\beta^{-2}\mu^\frac{1}{2}\Li_2(-e^{-\beta\mu})\ldots
\end{eqnarray}
The above expression for pressure is the most general analytic relation for the pressure of a degenerate Fermi gas at a finite temperature. 
The first term is the zero temperature pressure and the subsequent terms are the corrections due to the finite temperature of the gas, and include $\Li_s$, the polylogarithm functions of different orders $s$. The expression is terminated after the fourth term as the polylogs fall off exponentially as the gas becomes more and more degenerate. Eq. (\ref{pressure eqn3}) is a natural extension of the first-order Sommerfeld correction \citep{som28}. 

The central temperature of VLM stars and brown dwarfs are of the same order as that of the electron Fermi temperature and thus the degeneracy parameter $\psi$ is defined as
\begin{equation}\label{the degeneracy parameter}
\psi=\frac{k_BT}{\mu_F}=\frac{2m_ek_BT}{(3\pi^{2}\hbar^3)^{\frac{2}{3}}}\left[\frac{\mu_e}{\rho N_A}\right]^{\frac{2}{3}},
\end{equation}
where $\mu_F $ is the electron Fermi energy in the degenerate limit and $\frac{1}{\mu_e} = X +\frac{Y}{2}$ is the number of baryons per electron and $X$ and $Y$ are the  mass fractions of hydrogen and helium respectively. Other constants have their standard meaning.\\
Rewriting Eq. (\ref{pressure eqn3}) in terms of the degeneracy parameter $\psi$ and  retaining terms only up to second order, we arrive (for $\mu=\mu_F$) at 
\begin{eqnarray}\label{fermip}
P_F=\frac{2}{5}aA^{\frac{5}{2}}\left[\frac{\rho}{\mu_e}\right]^{\frac{5}{3}} \left[1-\frac{5}{16}\psi\ln(1+e^{-\frac{1}{\psi}}) \right. \\ \nonumber \left.  +\frac{15}{8}\psi^{2}\left\lbrace \frac{ \pi^2}{3}+\Li_2(-e^{-\frac{1}{\psi}}) \right\rbrace \right],
\end{eqnarray}
where $A=\frac{(3\pi^{2}\hbar^3N_A)^{\frac{2}{3}}}{2m_e}$ is a constant.
However, the interior of a brown dwarf is also composed of ionized hydrogen and helium. The total pressure is a combined effect of both electrons and ions, i.e., $P = P_F+P_{\rm ion}$,
where $P_F$ is the Fermi pressure for an ideal non-relativistic gas at a finite temperature. The pressure due to ions for an  ionized hydrogen gas can be approximated as 
\begin{equation}\label{ionp}
P_{\rm{ion}}=\frac{k\rho T}{\mu_1 m_H}.
\end{equation}

Therefore the final equation of state for the combined pressure is 
\begin{eqnarray}\label{the power law eq of state}
P=\frac{2}{5}aA^{\frac{5}{2}}\left[\frac{\rho}{\mu_e}\right]^{\frac{5}{3}}\left[1-\frac{5}{16}\psi\ln(1+e^{-\beta\mu}) \right. \\ \nonumber  \left. +\frac{15}{8}\psi^{2}\left\lbrace \frac{ \pi^2}{3}+\Li_2(-e^{-\frac{1}{\psi}}) \right\rbrace+\alpha\psi \right],
\end{eqnarray}
where $\alpha = \frac{5\mu_e}{2\mu_1}$ and $\mu_1$ is the mean molecular weight for helium and ionized hydrogen mixture and is expressed as
\begin{equation}
\frac{1}{\mu_1} = \left((1+x_{H^+})X+\frac{Y}{4}\right),
\end{equation}
where  $x_{H^+} $ is the ionization fraction of hydrogen. It should be noted that $x_{H^+} $ changes as one moves from the core (completely ionized) to the surface which is mainly composed of molecular hydrogen and helium. 

There are several corrections to the EOS that can be considered. The Coulomb pressure and the exchange pressure (see Eq. (13) in \cite{Ste91}) are two important corrections to Eq. (\ref{the power law eq of state}). However, as stated earlier they are less important for more massive brown dwarfs. \cite{Hub84} presents the contribution due to the electron correlation pressure, which depends on the logarithm of $r_e$, the mean distance between electrons. \cite{St096} present an analytic formulation of the EOS for fully ionized matter to study the thermodynamic properties of stellar interiors. They show that the inclusion of both electron and the ionic correlation pressure results in a $\sim 10\%$ correction to the EOS. Furthermore, \cite{Ger10} state that the main volume of the brown dwarfs and the interior of giant gas planets are in a warm dense matter state, where correlation energy, effective ionization energy and the electron Fermi energy are of the same order of magnitude, making it effectively a strongly correlated quantum system.  \cite{Beck15} give an EOS for hydrogen and helium covering a wide range of density and temperature. They extend their ab intio EOS to the strongly correlated quantum regime and connect it with the data derived using other methods for the neighboring regions of the $\rho - T$ plane. These simulations are within the framework of density functional theory molecular dynamics (DFT-MD) and give a detailed description of the internal structure of brown dwarfs and giant planets. This leads to a $2.5\%-5\%$ correction in the mass-radius relation. \\
The study of the EOS of brown dwarfs will help in understanding degenerate bodies in the thermodynamic regime that is not so close to the high pressure limit of a fully degenerate Fermi gas. In this context, the Mie-Grueneisen equation of state is of relevance to test the validity of the assumption that the Grueneisen parameter  $ \gamma = \left(\frac{\partial\,log \,T}{\partial\, log\,\rho}\right) _s $ is independent of the temperature $T$ \citep{And00} at a constant volume $V$. The brown dwarf regime is in a way more interesting than the white dwarf regime since it is not so close to the limit of a fully degenerate Fermi gas. In Appendix C we have provided analytic expressions for two parameters that are of particular relevance for the brown dwarfs; the specific heat ($C_v\;\rm {or}$ $C_p$) and the Grueneisen parameter.  


\section{An analytic model for brown dwarfs}
In this section, we derive some of the essential thermodynamic properties of a polytropic gas sphere based on the discussion in \cite{chand39}. As is evident from Eq. (\ref{the power law eq of state}), the $P-\rho$ relation for a brown dwarf is a polytrope
\begin{equation}
P=K\rho^{(1+\frac{1}{n})},
\end{equation}
where the index $n=3/2$. $K$  is a constant depending on the composition and degeneracy and can be expressed (from Eq. (\ref{the power law eq of state})) as
\begin{equation}\label{the K}
K= C \mu_{e}^{-5/3}\left(1+\gamma +\alpha\psi \right),
\end{equation}
where for a simplified presentation we represent the correction terms as 
\begin{equation}
\gamma=-\frac{5}{16}\psi\ln(1+e^{-\beta\mu})+\frac{15}{8}\psi^{2}\left\lbrace \frac{ \pi^2}{3}+\Li_2(-e^{-\frac{1}{\psi}}) \right\rbrace
\end{equation}
and $C\footnote {On using the values of natural constants we get \\ $C= 10^{13}\,\rm{cm^4g^{-2/3}s^{-2}}$ .}=\frac{2}{5}aA^{\frac{5}{2}}$ is a constant.
 The solution to the Lane-Emden equation subject to the zero pressure outer boundary condition can be used to arrive at useful results for  $R$, $\rho_c$ and $P_c$ for the polytropic equation of state, Eq. (\ref{the power law eq of state})
The radius can be expressed as 
\begin{equation}\label{radius1}
R=2.3573  \frac{K}{GM^{\frac{1}{3}}}
\end{equation}
\citep{chand39}. On substituting Eq. (\ref{the K}) for $K$,  the radius for a brown dwarf can be expressed as the function of degeneracy and mass, 
\begin{equation}\label{radius}
R=2.80858 \times 10^9\left(\frac{M_\odot}{M}\right)^{\frac{1}{3}} \mu_{e}^{-5/3}\left(1+\gamma
+\alpha\psi \right)\rm cm.
\end{equation}
Similarly, the expressions for the central density $\rho_c$ and central pressure  are given by the relations $\rho_c=\delta_n(\frac{3M}{4 \pi R^3})$ and $P_c=W_n\frac{GM^2}{R^4}$, where the constant $\delta_n = 5.991$ and $W_n=0.77$ for the polytrope of $n=1.5$ \citep{chand39}. On substituting the expression for $R$ (Eq. \ref{radius}) in these relations we get 

\begin{equation}\label{the central density}
\rho_c=1.28412\times 10^{5}\left(\frac{M}{M_\odot}\right)^{2}\frac{\mu_e^5}{(1+\gamma+\alpha\psi )^3}\rm g/cm^3
\end{equation}

and 

\begin{equation}
P_c=3.26763\times 10^{9}\left(\frac{M}{M_\odot}\right)^{\frac{10}{3}}\frac{\mu_e^{\frac{20}{3}}}{(1+\gamma+\alpha\psi )^4} \rm Mbar.
\end{equation}

These are the scaling laws of the density and pressure in the interior core of a brown dwarf. Interestingly, these vary with the degeneracy parameter $\psi$ that 
is a function of time. Thus a very simple polytropic model can yield the time evolution of the internal thermodynamical conditions of a brown dwarf. From the definition of the degeneracy parameter in Eq. (\ref{the degeneracy parameter}) and using Eq. (\ref{the central density}), the central temperature can be expressed as a function of $\psi$:
\begin{equation} \label{the central temperature}
T_c = 7.68097\times 10^{8} \textrm{K} \left(\frac{M}{M_\odot}\right)^{\frac{4}{3}}\frac{\psi \, \mu_e^{\frac{8}{3}}}{(1+\gamma+\alpha\psi )^2}.
\end{equation}
The central temperature has a maximum for a certain value of $\psi$, and it increases for greater values of $M$. Further, using Eq. (\ref{radius}) we have shown the variation of central temperature $T_c$ as a function of radius $R$.  $T_c$ increases as the object contracts under the influence of gravity. It peaks at a certain $R$ and then cools over time. The maximum peak temperature increases for heavier objects and also depends on the extent of ionization of hydrogen and helium. Figure 1 shows the variation of the central temperature as a function of radius for different mass ranges. If the critical temperature for thermonuclear reactions is around $3 \times 10^6 \, \rm K $, we can roughly  estimate the critical mass for the main sequence as $ \sim 0.085\,M_\odot $. This is  similar to the estimated critical mass ($\sim 0.084\,M_\odot$) for the main sequence (see Figure 1 in \cite{Ste91}). However it should be noted that the estimate of minimum mass is very sensitive to the mean molecular mass $\mu_1$. In the Fig. 1 we have used $\mu_1= 1.23$ for fully neutral gas (similar to $\bar{A}\sim 1.24$ for cosmic mixture as used in \cite{Ste91}). Depending upon the value of $\mu_1$ the minimum mass may vary significantly. For example, if we consider a fully ionised gas, i.e., $\mu_1=0.59$, it yields a minimum mass of $0.12\,M_\odot $.  \\  



\begin{figure}
\centering
\includegraphics{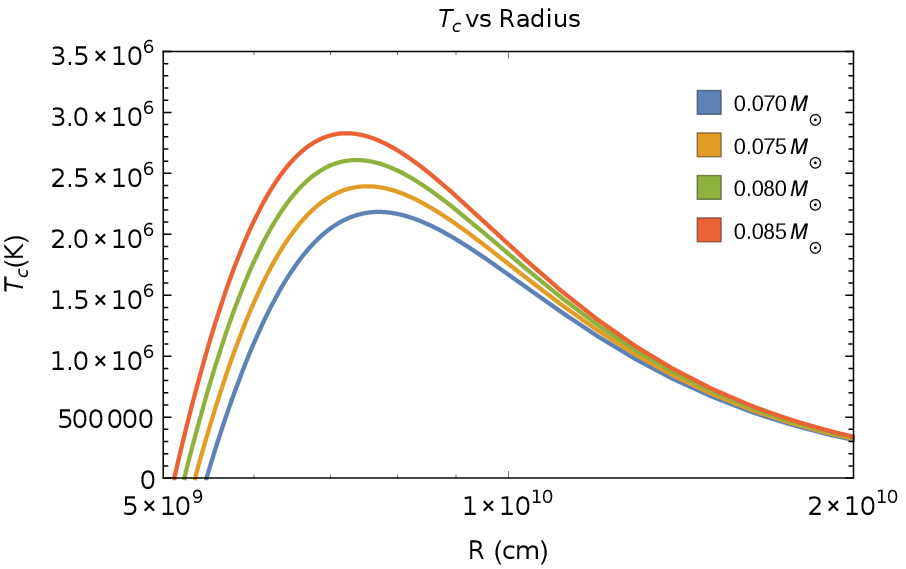}
\figcaption{The variation of $T_c$ versus radius $R$  for different masses.}
\end{figure}

\section{Surface properties }
In this section we discuss a very simple but crude model which is broadly based on the phase transition proposed by \citet{Chab92} and the isentropic nature of the interior of brown dwarfs. The development of a theoretical model for studying the variation of surface luminosity over time for low mass stars (LMS) and brown dwarfs is a great challenge. There is no stable phase of nuclear burning for brown dwarfs and the luminosity gradually decreases with time. This leads to an age-mass degeneracy in observational determinations. Our lack of knowledge in understanding the physics of the interior of brown dwarfs restricts the development of a comprehensive model. However, extensive simulations were done on the molecular-metallic  transition of hydrogen for LMS and planets  \citep{Chab92,Chab& horn92}. The \cite{Chab92} model predicts a first order transition for the metallization of hydrogen at a pressure of $\sim 1 \,\rm Mbar$ and critical temperature of $\sim 15300\, \rm K$. Such pressure and temperature values are appropriate for giant planets and brown dwarfs. Modern numerical simulations \citep{Yan15, Mor10} do confirm the existence of such phase transitions at the same pressure range but predict a much different range of temperature $\sim 2000\,\textrm{K} -3000 \, \rm K$. This new temperature regime is certainly too low for brown dwarfs. Although the pressure estimate is relatively well established in these numerical simulations, the phase transition temperature is still a matter of continuing investigation \citep{Beck15}.
Having noted these caveats, we present the existing model for the surface temperature, based on the \citet{Chab92} and \cite{Burs93}. We also introduce a simpler treatment of the specific entropy.

The PPT occurs over a narrow range of densities near $1.0\, \rm{g/cm}^3$ from a partially ionized phase ($x_{H^+} \sim 0.5$) to a neutral molecular phase ($x_{H^+} < 10^{-3})$. For massive brown dwarfs the phase transition occurs nearer to the surface. \cite{Burs93} used the following approximate analytic expressions for the specific entropy (Eqs. (2.48) and (2.49) in \cite{Burs93}) for the two phases of the PPT:

\begin{equation}
\sigma_1 = -1.594 \ln \frac{1}{\psi} +12.43,
\end{equation}

\begin{equation}
\sigma_2 = 1.032 \ln T/\rho^{0.42}-2.438.
\end{equation}
Similar expressions for the entropy at the interior and the atmosphere are given in Eqs. (21) and (22) in \cite{Ste91}. We use a simplified approach to  make the origin of the above equation clear in the spirit of an analytic model. We derive analytic expressions for the entropy of the ionized and the molecular hydrogen separately for the two phases (similar to equations (2.48) and (2.49) in \cite{Burs93}) and match them via the phase transition.}  It is assumed that the presence of helium does not affect the hydrogen PPT \citep{Chab92}.

The region between the strongly correlated quantum regime and the ideal gas limit can be modelled with corrections to the ideal gas equation. Such correction terms can be expressed by virial coefficients (see Eq. (1) in \cite{Beck15}).  
For simplicity, we ignore such corrections in our EOS (Eq. (\ref{the power law eq of state})) and consider only the contribution of electron pressure, Eq.(\ref{fermip}), and the ion pressure, Eq. (\ref{ionp}), of the partially ionized hydrogen (of ionization fraction $ x_{H^{+}} $) and helium mixture.

The total entropy for our EOS (Eq. (\ref{the power law eq of state})) is the sum of the entropies of the atomic/molecular gas  and the degenerate electrons. The internal energy per gram for the monoatomic gas particles is
\begin{equation}\label{internal energy}
U= \frac{3}{2}\frac{k_B N_0 T}{ \mu_1}.
\end{equation}
Ideally, we can consider the total energy as a combination of kinetic energy, radiation energy and the ionization energy (Eq. \ref{the total energy}). But as the electron gas is degenerate, the radiation pressure is relatively unimportant.  Furthermore, as shown in Appendix B, we may even neglect the contribution of the ionization energy as the gas is only partially ionized. Taking the partial derivatives of the expression for the internal energy Eq. (\ref{internal energy}) we can express the change in heat $dQ$ using the first law of thermodynamics, Eq. (\ref{1st law}). However, as the ionization fraction $x_{H^{+}}(\rho,T)$ is a function of density and temperature we further use Saha's ionization equations (Eqs. \ref{saha1}, \ref{saha2}) to get
\\

\begin{eqnarray}\label{dq}
\frac{dQ}{T} = -\frac{3}{2} \frac{k_B N_0}{\mu_1}\frac{dT}{T}+ \frac{k_B N_0}{\mu_1} \frac{dW}{W} + \\ \nonumber \left(\frac{3}{2}\right)^2 Hk_B N_0\frac{dT}{T}+ \frac{3}{2} Hk_B N_0\frac{dV}{V},
\end{eqnarray}
where $W=T^3V$, $H=\frac{x_{H^{+}}(1-x_{H^{+}})}{2-x_{H^{+}}}$. A more generalized version including the radiation and the ionization terms are shown in Eq. (\ref{general entropy appn}) in Appendix B. For $ TdS =dQ $ we integrate Eq. (\ref{dq}) to express the entropy in the interior of the brown dwarf as
\begin{equation}\label{the interior entropy}
S_1 = \frac{k_B N_0}{\mu_{1\rm{mod}}} \ln \frac{T^\frac{3}{2}}{\rho_1} + C_1,
\end{equation}
where 
\begin{equation}
\frac{1}{\mu_{1\rm{mod}}}=\left(\frac{1}{\mu_1}+\frac{3}{2}\frac{x_{H^{+}}(1-x_{H^{+}})}{2-x_{H^{+}}}\right), 
\end{equation}
and $\mu_1$ is different for each model in this region and is calculated using $x_{H^{+}}$ from Table 1. However, the contributions to entropy due to radiation and the degenerate electrons (see Eq 2-145 in \cite{Clay68}) are negligible  in the range of temperature and density applicable for brown dwarfs.
Based on a similar argument, the analytic expression for the entropy of non-ionized molecular hydrogen and helium mixture at the photosphere is expressed as 
\begin{equation}\label{the exterior entropy}
S_2 = \frac{k_B N_0}{\mu_2} \ln \frac{T^\frac{5}{2}}{\rho_2} + C_2,
\end{equation}
where $\frac{1}{\mu_2} = \frac{X}{2}+\frac{Y}{4}$ is the mean molecular weight for the hydrogen and helium mixture. The expression for entropy $S_2$ is derived using the first law of thermodynamics (Appendix B) and the relation of the internal energy of diatomic molecules ($ U=\frac{5}{2} \frac{k_BN_0T}{\mu_2}$). Here we have considered only five degrees of freedom as the temperatures are just sufficient to excite the rotational degrees of H$_2$ but the vibrational degrees remain dormant. It should be noted that Eqs. (\ref{the interior entropy}) and (\ref{the exterior entropy}) are just simplified forms of the entropy expressions presented in \cite{Burs93} and \cite{Ste91}.

 Thus the entropy in the two phases is dominated by contributions from the ionic and molecular gas, respectively.
Using the same argument as \cite{Burs93}, that the two regions of different temperature and density are separated by a phase transition of order one we can estimate the surface temperature.
Using the expression for the degeneracy parameter $\psi$ from Eq. (\ref{the degeneracy parameter}) we can simplify Eq. (\ref{the interior entropy}) to be
\begin{equation}\label{entropy1mod}
S_1 = \frac{3}{2}\frac{k_B N_0}{\mu_{1\rm{mod}}}( \ln \psi +12.7065) + C_1.
\end{equation}
Furthermore, the jump of entropy 
\begin{equation}\label{entropyjump}
\Delta \sigma=\frac{S_2-S_1}{k_BN_0},
\end{equation}
(see Table 1) for the phase transition at each point of the coexistence curve of PPT \citep{Saumon89}, is used to estimate the relation  $\mid C_1-C_2\mid$ between the two constants in Eqs. (\ref{the interior entropy}) and (\ref{the exterior entropy}). For $T=T_{\rm{eff}}$ and $\rho_2=\rho_e$ in Eq (\ref{the exterior entropy}), we can use Eqs. (\ref{entropy1mod}) and (\ref{the exterior entropy}) in Eq. (\ref{entropyjump}) and the value of $\mid C_1-C_2\mid$ to obtain a wide range of possible values of surface temperature $T_{\rm{eff}}$ in terms of the degeneracy parameter and photospheric density $\rho_e$:
\begin{equation}\label{effective temperature}
T_{\rm{eff}}= b_1\times 10^{6} \rho_e^{0.4} \psi^v \:\:\rm{K}.
\end{equation}
The values of the parameters $b_1$ and $v$ for different models are shown in Table \ref{tbl:model}.   According to the Chabrier model, the  critical temperature $ \sim 1.53 \times 10^4$ K and critical density $ \sim 0.35 $ g cm$^{-3}$ marks the end of the phase transition with $ \Delta \sigma=\frac{\Delta S}{k_BN}=0 $.   In the following discussion we briefly summarise the steps from \cite{Burs93}. We replace Eq. (2.50) in \cite{Burs93} by Eq. (\ref{effective temperature})  to estimate  the surface luminosity.  As an example we select a particular phase transition point (model D) and show the derivation of surface luminosity using hydrostatic equilibrium and the ideal gas approximation. The photosphere of a brown dwarf is located at approximately the $\tau = \frac{2}{3}$ surface, where 
\begin{equation}\label{Tau}
\tau=\int_r^{\infty} \kappa_R \rho \textit{ dr}.
\end{equation}
is the optical depth.
Using the general equation for hydrostatic equilibrium, $dP=-(GM/r^2)\rho \, \textit{dr}$, and Eq. (\ref{Tau}), the photospheric pressure can be expressed as 
\begin{equation}\label{external pressure 1}
P_e=\frac{2}{3}\frac{GM}{\kappa_R R^2},
\end{equation}   
where $\kappa_R $ is the Rosseland mean opacity and the other variables have their standard meanings.
Furthermore, our EOS (Eq. (\ref{the power law eq of state})) in the approximation of negligible degeneracy pressure near the photosphere gives the photoshperic pressure as
\begin{equation}\label{external pressure 5}
P_e=\frac{\rho_e N_Ak_B T_{\rm{eff}}}{\mu_2}.
\end{equation}
Now using the expression for radius $R$ (Eq. \ref{radius}) in Eq. (\ref{external pressure 1}), we can calculate the external pressure $P_e$ as a function of $M$ and $\psi$:
\begin{equation}\label{external pressure}
P_e= \frac{11.2193 \,\rm bar}{\kappa_R}\left(\frac{M}{M_{\odot}}\right)^{5/3} \frac{\mu_e^{10/3}}{(1+\gamma
+\alpha\psi)^2} .
\end{equation}
On using Eq. (\ref{external pressure}) in Eq. (\ref{external pressure 5}) and substituting  $T_{\rm{eff}}$ for model D with $b_1=2.00 $ and $v=1.60$ from Table \ref{tbl:model}, the effective density $\rho_e$ can be expressed as a  function of $M$ and $\psi$:
\begin{equation}\label{the effective density}
\rho_e^{1.40}=\frac{6.89811}{\kappa_R N_A k_B} \left(\frac{M}{M_{\odot}}\right)^{5/3} \frac{\mu_e^{10/3}\mu_2 \; \rm {g/cm^3} }{(1+\gamma+\alpha\psi)^{2}\psi^{1.58}}.
\end{equation}
Substituting the expression for $\rho_e$ from Eq. (\ref{the effective density}) in Eq. (\ref{effective temperature}) we derive the expression for effective temperature for model D as a function of $M$ and $\psi$:
\begin{equation}\label{t effective}
T_{\rm{eff}}=\frac{2.57881 \times 10^4 \, \rm K }{\kappa_R^{.2856}} \left(\frac{M}{M_{\odot}}\right)^{0.4764} \frac{\psi^{1.1456}}{(1+\gamma+\alpha\psi)^{0.5712}} .
\end{equation}
Similarly, the surface temperature can be evaluated for all the other models. Since the procedure is same for all the models in Table \ref{tbl:model}, we just show one calculation. For this range of surface temperatures, the Stefan-Boltzmann law, $L=4\pi R^2 \sigma T_{\rm{eff}}^4$, yields a set of possible values of the surface luminosity $L$ as a function of the degeneracy parameter $\psi$. The luminosity for model D using Eq. (\ref{radius}) and Eq. (\ref{t effective}) is
\begin{equation}\label{luminosity model D}
L=\frac{0.41470\times L_{\odot}}{\kappa_R ^{1.1424}} \left(\frac{M}{M_{\odot}}\right)^{1.239} \frac{\psi^{4.5797}}{(1+\gamma+\alpha\psi)^{0.2848}},
\end{equation}
where $\sigma$ is the Stefan-Boltzmann constant.
Substellar objects below the main sequence mass gradually evolve towards complete degeneracy and a state of stable equilibrium as their luminosity decreases over time. In the following sections we show that the degeneracy parameter $\psi$ is a function of time and it evolves toward $\psi =0$ over the lifetime of brown dwarfs. This gives us an estimate of the luminosity at different epochs of time. 

\begin{deluxetable}{lcccccccc}
\tabletypesize{\footnotesize} \tablewidth{-4pt} 
\tablecaption{Effective temperature for different phase transition points\tablenotemark{a}  \label{tbl:model}} 
\tablehead{
\colhead{${\rm Model} $} & 
\colhead{$\log T$ (K)} &
\colhead{$P$(Mbar)} & 
\colhead{$\rho_1$ (g cm$^{-3}$)} &
\colhead{$\rho_2$ (g cm$^{-3}$)} &
\colhead{$\Delta \sigma$ } &
\colhead{$2x_{H^{+}}$ } &
\colhead{$b_1 $} &
\colhead{$v$} }
\startdata 
A &3.70 & 2.14      &  0.75 & 0.92 & 0.62 & 0.48 & 2.87 & 1.58   \\
B &3.78 & 1.95        &  0.70 & 0.88 & 0.59 & 0.50 &2.70 & 1.59  \\
C &3.86 & 1.62      &  0.64 & 0.80 & 0.54 & 0.50 & 2.26 & 1.59  \\
D &3.94 & 1.39 & 0.58 & 0.74 & 0.51 & 0.51 & 2.00  & 1.60  \\
E &4.02 & 1.13     &  0.51 & 0.65 & 0.46 & 0.52 &1.68  & 1.61 \\ 
F &4.10 & 0.895       &  0.43 & 0.55 & 0.42& 0.50& 1.29  & 1.59 \\
G &4.18 & 0.631       &  0.35 & 0.38 & 0.14& 0.33&0.60  & 1.44  \\
H &4.185 & 0.614     &  0.35 & 0.35 & 0.00& 0.18& 0.40  & 1.30 \\
\enddata
\tablenotetext{a}{The phase transition points are taken from \citet{Chab92}. This gives the possible range of surface temperature depending on the phase transition points. For different values of temperature and density at which the phase transition takes place, the effective surface temperature is calculated using Eq. (\ref{effective temperature}).}
\end{deluxetable}

\subsection{Validity of PPT in brown dwarfs}
There is a distinction between the temperature-driven PPT with a critical point at $ \sim 0.5$ Mbar and between $10000$ K and $20000$ K as predicted by the chemical models \citep{Saumon95}, and the pressure-driven transition from an insulating molecular liquid to a metallic liquid with a critical point below 2000 K at pressures between $1$ and $3$ Mbars. The latter is predicted e.g., by \cite{Lor09}, \cite{Maz10}, and \cite{Mor10} based on the ab initio simulations . \cite{Lor09} rule out the presence of PPT above $10000$ K and give an estimate of the critical points for the transition at $T_c = (1400 \pm 100)\, \rm K$, $P_c = 1.32 \pm 0.1$ Mbar, $\rho_c = 0.79 \pm 0.05$ $\rm g/cm^3$. Similarly \cite{Mor10} estimated the critical point of the transition at a temperature near $2000$ K and pressure near $1.2$ Mbar. Signatures of pressure-driven PPT in a cold regime below $2000$ K are obtained by \cite{Knu15}. Figure 1 in \cite{Knu15} shows the melting line (black) as well as the different predictions for the coexistence lines for the first order transition (green curves).  Brown dwarf interior temperatures are far above these estimates for a first order transition from the insulating to the metallic system. The same is true for Jupiter. Of course a continuous transition may be possible in Jupiter and brown dwarfs, but a first order transition may not be possible. Thus the determination of the range of temperature of this transition  provides a much needed benchmark for the theory of the standard models for the internal structures of the gas-giant planets and low mass stars.  

\section{Nuclear processes}

VLM stars and brown dwarfs contract during their evolution due to gravitational collapse. The core temperature increases and the contraction is halted by either the degeneracy pressure of the electrons or the onset of the nuclear burning, whichever comes first. In the first case, the brown dwarf continues to lose energy through radiation and cools down with time without any further compression. However, massive brown dwarfs or stars at the edge of the main sequence can burn hydrogen for a very long time before they either cease nuclear burning or settle into a steady state main sequence.
The thermonuclear reactions suitable for the brown dwarfs and VLM stars are 

\begin{equation}\label{deuterium formation}
p + p \rightarrow d+e^++\nu,
\end{equation}
\begin{equation}\label{deuterium burning}
	p+d\rightarrow  {}^{3}{\rm He}+\gamma.
\end{equation}

As the central temperature is not high enough to overcome the Coulomb barrier of the $ \rm{{}^3He-{}^3He}$ reaction and the p-p chain is truncated, $\rm{{}^4He} $ is not produced. Most of the thermonuclear energy is produced from the burning of the primordial deuterium, Eq. (\ref{deuterium burning}). The energy generation rates of the above processes are given as 
\begin{equation}
\dot{\epsilon}_{pp}=2.5\times10^{6}\left[\frac{\rho X^2}{T_6^{2/3}}\right] e^{\frac{-33.8}{T_6^{1/3}}}\rm{erg/g \cdot s} \, ,
\end{equation}
\begin{equation}
\dot{\epsilon}_{pd}=1.4\times10^{24}\left[\frac{\rho XY_d}{T_6^{2/3}}\right] e^{\frac{-37.2}{T_6^{1/3}}}\rm{erg/g \cdot s} \, ,
\end{equation}
\citep{Fow75}. However one can fit the thermonuclear rates to a power law in $T$ and $\rho$ in terms of the central temperature ($T_c$) and density ($\rho_c$) as in \citet{Fowler64}:
\begin{equation}\label{power law}
\dot{\epsilon_n}= \dot{\epsilon_c}\left[\frac{T}{T_c}\right]^s \left[ \frac{\rho}{\rho_c}\right]^{u-1} ,
\end{equation}
where $u \simeq 2.28$ and $s=6.31$ are constants that depend on the core conditions \citep{Burs93}.
To obtain the luminosity due to the nuclear burning $L_N = \int \dot{\epsilon_n}dm$, we use the power law form for the nuclear burning rate $\dot{\epsilon}_{n}$, (Eq. \ref{power law}), and making the polytropic approximation $\rho = \rho_c\theta ^n$ and setting  $\frac{T}{T_c}=\left(\frac{\rho}{\rho_c}\right)^{2/3}$, we obtain 
\begin{equation}\label{ln}
L_N=\int \dot{\epsilon_n}dm=4\pi a^3\dot{\epsilon_c}\rho_c\int\theta^{n(u+2s/3)}\zeta^2d\zeta.
\end{equation}
where $r=a \zeta$ \citep{chand39}.
Inserting Eqs. (\ref{the central density}) and (\ref{the central temperature}) in Eq. (\ref{ln}) yields the final expression for luminosity as 

\begin{equation}\label{nuclear luminosity}
L_N= 7.33\times 10^{16} L_{\odot} \left(\frac{M}{M_{\odot}}\right)^{11.977} \frac{ \psi^{6.0316}}{(1+\gamma+\alpha\psi)^{16.466}}.
\end{equation}

\section{Estimate of the minimum mass}

In this section we estimate the minimum main sequence mass by comparing the surface luminosity (Eq. \ref{luminosity model D}), with the luminosity ($L_N$) due to nuclear burning at the core of LMS and brown dwarfs. Instead of just quoting one value as the critical mass, we have presented a range of values depending on the various phase transition points listed in Table 1. This will give us a range of values for the minimum critical mass that is sufficient to ignite hydrogen burning.
 Model H marks the end of the phase transition and gives a lower limit of the critical mass.  However, we calculate the mass limit for model B and  model D only.
Equating $L_N$ of Eq. (\ref{nuclear luminosity}) with $L$ of Eq. (\ref{luminosity model D}) gives us 
\begin{equation}\label{the critical mass equation}
\frac{M}{M_\odot}=\frac{0.02440}{\kappa_R^{0.106}} \frac{(1+\gamma+\alpha \psi)^{1.507}}{\psi^{0.1617}}=F(\psi),
\end{equation}
where $\alpha =2.32$ for $\mu_e=1.143$ and $\mu_1=1.24$, which are the number of baryons per electron and the mean molecular weight of neutral ($x_{H^+}=0$) hydrogen and helium, respectively. These mass densities are evaluated for hydrogen and helium mass fractions of $X=0.75$ and $Y=0.25$, respectively.\\
The right hand side of Eq. (\ref{the critical mass equation}) has a minimum at a certain value of $\psi$. This gives the lowest mass at which Eq. (\ref{the critical mass equation}) has a solution and this corresponds to  the boundary of brown dwarfs and VLM stars. The minimum of $F(\psi)$ is at $\psi_{min}=0.042$. Substituting this in Eq. (\ref{the critical mass equation}) and for $\kappa_R=0.01$ $\rm{cm^2/g}$, the minimum mass (model D) is
\begin{equation}
M=0.078\, M_\odot.
\end{equation}
A similar analysis for model B gives the value of minimum mass of $M=0.085M_\odot$ for 
$\psi_{min}=0.042$. For the other models the minimum main sequence mass is in the range of $0.064-0.087M_\odot$.\\
The solution is relatively independent of mean molecular weight $\mu_1 $. For example, using partially ionised gas i.e. $\mu_1=0.84$ in $\alpha$, it increases the minimum stellar mass by only $\sim 5\%$.
\\

\section{A cooling model}
A simple cooling model for a brown dwarf is presented in both  \cite{Burs93} and \cite{Ste91}. In this section we review some of these steps using our more exact EOS Eq. (\ref{the power law eq of state}) and represent the evolution of the brown dwarfs over time. Using the first and the second law of thermodynamics, the time varying energy equation for a contracting star is expressed as
\begin{equation}
\frac{dE}{dt}+P\frac{dV}{dt}=T\frac{dS}{dt}=\dot{\epsilon}-\frac{\partial L}{\partial M},
\end{equation}
where $S$ is the entropy per unit mass and the other symbols have their standard meaning. The energy generation term $\dot{\epsilon}$ is ignored. On integrating over mass we get,
\begin{equation}\label{Energy integral}
\frac{d\sigma}{dt}\left[\int N_Ak_BTdM\right]= -L,
\end{equation}
where $L$ is the surface luminosity and $\sigma =\frac{S}{k_BN_A}$. Now replacing $T$ in terms of the degeneracy parameter $\psi$ in Eq. (\ref{the degeneracy parameter}), and using the polytropic relation $P=K\rho^{\frac{5}{3}}$, we arrive at
\begin{equation}\label{the rate of entropy}
\frac{d\sigma}{dt} \frac{N_Ak_B \psi}{\mu_e^{\frac{2}{3}}}\int PdV = -L.
\end{equation}
Using the standard expression, $\int PdV= \frac{2}{7}\frac{GM^2}{R}$, for polytropes of $n=1.5$, the integral in  Eq. (\ref{Energy integral}) reduces to 
\begin{equation}\label{the mass equation}
\int N_Ak_BTdM =  \frac{6.73857 \times 10^{49}\,\psi \mu_e^{\frac{8}{3}}}{(1+\gamma+\alpha\psi)^2}\left(\frac{M}{M_\odot}\right)^{\frac{7}{3}}.
\end{equation}
The variation of the entropy with time (Eq. \ref{Energy integral}) can be expressed as the rate of change of degeneracy over time.  As the star collapses, the gas in the interior becomes more and more degenerate and finally the degeneracy pressure halts further contraction. A completely degenerate star $(\psi = 0)$ becomes static and cools with time. 
Thus by substituting the time variation of entropy, using Eq. (\ref{entropy1mod}), i.e 
\begin{equation}
\frac{d\sigma}{dt}= \frac{1.5}{\mu_{1mod}}\frac{1}{\psi} \frac{d \psi}{dt},
\end{equation}
and Eq. (\ref{the mass equation}) into the energy equation Eq. (\ref{Energy integral}), and using the luminosity expression for model D (Eq. \ref{luminosity model D}), we obtain an evolutionary equation for $\psi$:

\begin{eqnarray}\label{the differential equation}
\frac{d\psi}{dt}= \frac{9.4486 \times 10^{-18}}{\kappa_R ^{1.1424}} \left(\frac{M_{\odot}}{M}\right)^{1.094} \times \\ \nonumber (1+\gamma+\alpha\psi)^{1.715} \psi^{4.5797}.
\end{eqnarray}
This is a nonlinear differential equation of $\psi$ for the model D, and an exact solution can only be obtained numerically. However we use some very simple and physical approximations to solve this differential equation to yield a simple relation of $\psi$ as a function of time and mass $M$. As we are trying to estimate the critical mass for the hydrogen burning, it is safe to ignore the early evolution of VLM stars and brown dwarfs. Thus we will solve the differential equation (\ref{the differential equation}) with the assumption  $\psi \ll 1$. Thus we can drop the term $(1+\gamma+\alpha\psi)^{1.715}$ as it is almost unity in the range $0 <\psi<0.1$ and integrate Eq. (\ref{the differential equation}) in the above limit to obtain 
\begin{equation}\label{psi with t}
\psi = \left(317.8+2.053 \times 10^{-6} \left(\frac{M_\odot}{M}\right)^{1.094} \frac{t}{\rm yr}\right)^{-0.2794}.
\end{equation}
Similarly, we can solve for $\psi$ for all other models and obtain the evolution of degeneracy over time. We  use this expression of $\psi$, for model D, and can express luminosity as a function of time $t$ and mass $M$. The time evolution of luminosity (model D) is represented in Figure 2. It is evident that such low mass objects continue to have low luminosity for millions of years before they gradually start to cool. For $t > 10^{7} $ yr, the luminosity declines as a function of time, $L \simeq t^{-1.2}$, as shown in Figure 2 (dashed black line). A simplified expression of the variation of luminosity after $10^7$ yr for model D is
\begin{equation}
L \simeq  L_{\odot} \left( \frac{M}{M_{\odot}}\right)^{2.63} \left(\frac{t}{10^7 \rm{yr}}\right)^{-1.2}.
\end{equation} 
Our luminosity model is consistent with the simulation results of the present day stellar evolution code Modules for Experiments in Stellar Astrophysics (MESA) (see figure 17 in \cite{Bill11}). \cite{Bill11} use a one-dimensional stellar evolution module, MESA star, to study evolutionary phases of  brown dwarfs, pre-main-sequence stars, and LMS.

\begin{figure}\label{f1}
\centering
\includegraphics{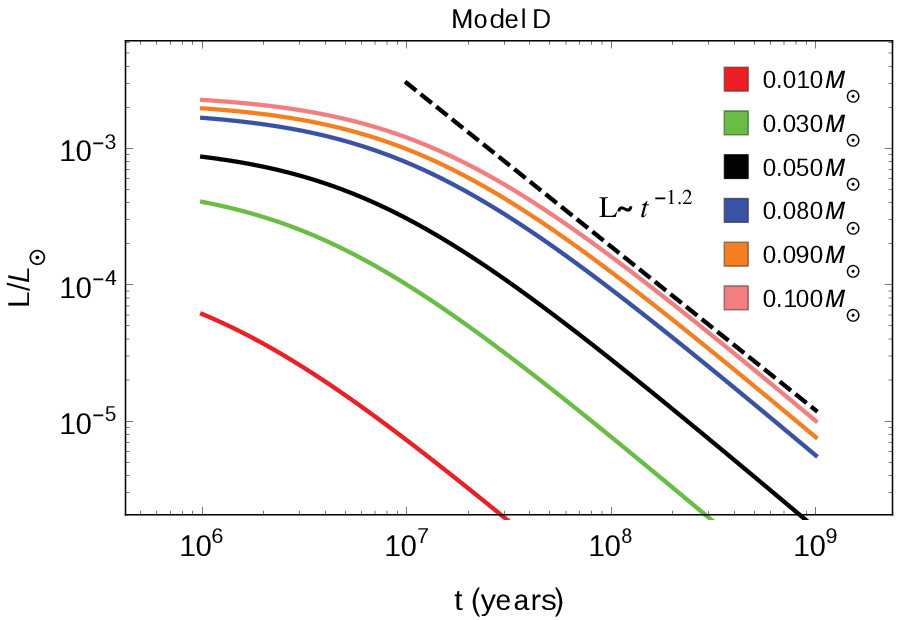}
\figcaption{The variation of $L/L_\odot$ over time $t$ for different masses.}
\end{figure}

In Figs. 3 and 4 the ratio $L_N / L$ is plotted against time for different masses in the substellar regime for models B and D, respectively. As evident for both the models there is a non-steady state of substantial nuclear burning for millions of years for substellar objects. For a critical mass of $0.085M_{\odot}$ (model B) and $0.078M_{\odot}$ (model D), the ratio $L_N/L$ approaches 1 in about a few billion years and marks the beginning of main sequence nuclear burning \footnote{Note that the time to reach the main sequence will increase if we use a partially ionized gas. For example, it becomes $\approx $10 Gyr if we use $\mu_1=0.84$.}. Stars with greater mass reach a steady state where the thermal energy balances the gravitational collapse. However, as our model does not consider any feedback from hydrogen fusion, the curves do not stabilize to a steady state main sequence regime, in which $L_N/L$ remains 1 until nuclear burning stops. Interestingly, the ratio $L_N/L$ is close to unity for many objects below the main sequence transition mass. This suggests that they burn nuclear fuel for a part of their evolutionary cycle but do not have enough mass to sustain a steady state.\\
Note that the results of \cite{Beck15} discussed in Section 2 would affect the luminosity $L$ by less than a few percent. For example if the radius $R$ increases (or decreases) by $2.5 \, \%$  for a constant value of mass $M$, $K$ in Eq. (\ref{radius1}) increases by $2.5\,\%$ and $T_{\rm eff}$ (Eq. \ref{t effective}) decreases by $1.5 \, \%$, therefore $L$ decreases by $\sim 1\,\%$.

\subsection{Brown Dwarfs as clocks}

 Interestingly the cooling properties of brown dwarfs (Fig. 2) can be calibrated to serve as an astronomical clock. As the electron degeneracy pressure puts a lower limit to the size of the dwarf, it cools slowly and radiates its internal energy. The luminosity of a brown dwarf is the most directly accessible observable quantity. As luminosity is a time variable, one can get important information on the age of a brown dwarf depending on its mass and the cooling rate. As evident from Figure 2, given mass and the luminosity one can roughly identify the age of the dwarfs. However it is still a challenge to estimate the mass of a brown dwarf. An essential part of the solution is to find brown dwarfs in a binary system where one can get an accurate estimate of the mass and then compare its luminosity against available models. Newly discovered brown dwarfs in eclipsing binaries \citep{Sta06,Tre15} can provide a data set of directly measured mass and radii. This can yield an empirical mass-radius relation that also tests the prediction of the theoretical models.
Furthermore, lithium in brown dwarfs has been used as a clock to obtain the ages of young open clusters as originally suggested by \cite {Mat94} and \cite{Bas96}, and most recently applied to the Pleiades by \cite{Dah15}. Massive brown dwarfs ($M >  0.065 M_{\odot}$) deplete their lithium on a longer time scale, but VLM stars and objects above the hydrogen burning limit fuse lithium on a much shorter time scale \citep{Mag93}.
A limitation of our model is that it does not include rotation. But the mechanical equilibrium in our models may not be significantly affected by this. For example, in model D, using  Eq. (\ref{psi with t}) in Eq. (\ref{radius}) we find the radius of a $10^7$ yr old brown dwarf of mass  $0.075 M_{\odot}$ to be $\sim 8 \times 10^9$cm. This implies that for a median observed rotational period ($2\pi/\omega$) of one day \citep{sch15}  the ratio of magnitude of the rotational energy ($\sim MR^2\omega^2$) to gravitational energy($\sim GM^2/R$) for a $0.075 M_{\odot}$ brown dwarf is $\sim 10^{-4}$.  However, convective mixing and the consequent lithium abundance has a strong connection to the rotation rate of brown dwarfs and pre-main-sequence stars \citep{Matc96}. Fast rotators are  lithium-rich compared to their slow rotating counterparts, indicating a connection between the lithium content and the spin rate of young pre-main-sequence stars and brown dwarfs \citep{Bou16}. Rapid rotation reduces the convective mixing, resulting in a higher lithium abundance in fast rotating  pre-main-sequence stars. Thus, the rotational evolution of a brown dwarf can potentially be used as a clock as discussed in \cite{sch15}.    
   

\begin{figure}\label{f2}
\centering
\includegraphics{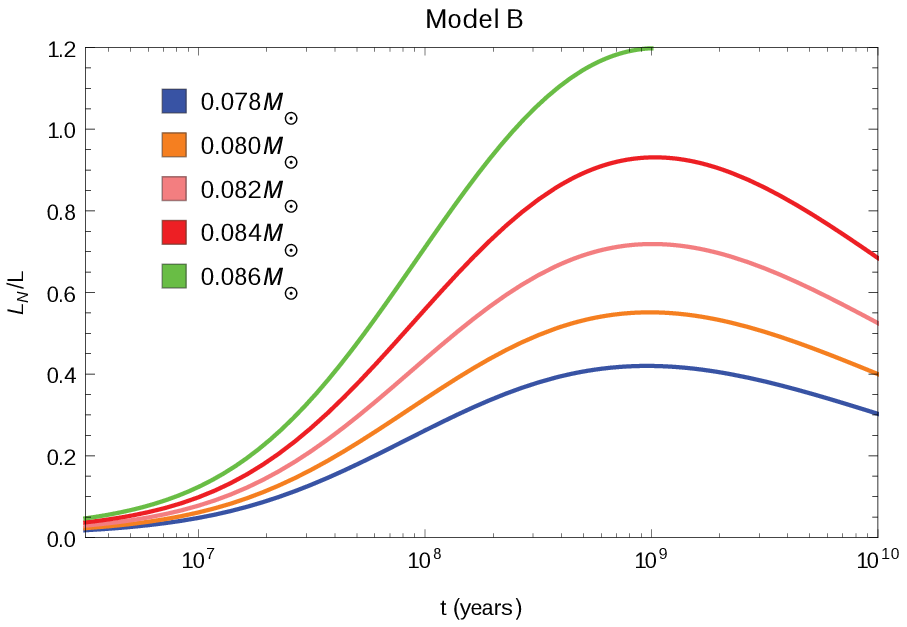}
\figcaption{The variation of $L_N/L$ over time $t$ for different  masses for model B in Table 1.}
\end{figure}

\begin{figure}\label{f3}
\centering
\includegraphics{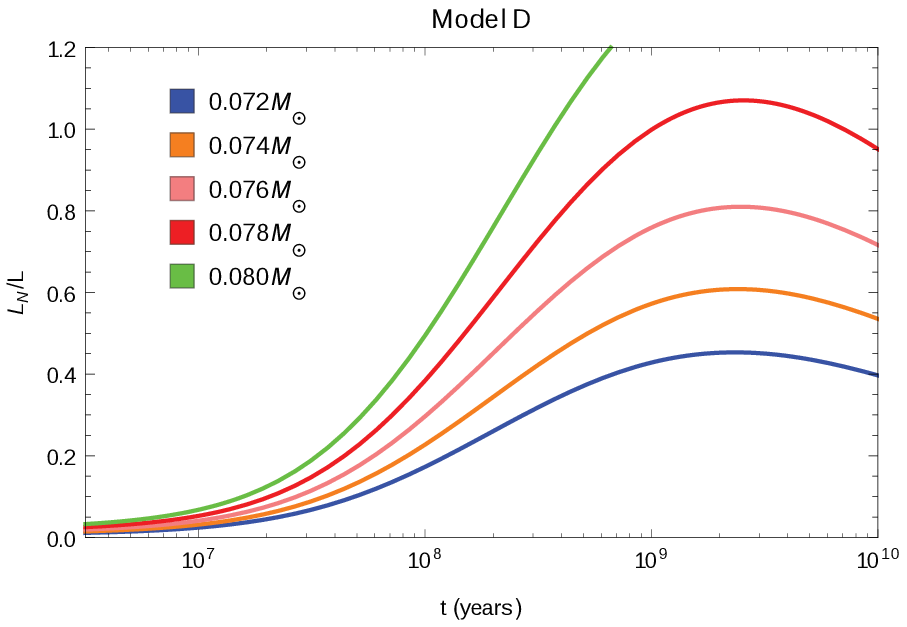}
\figcaption{The variation of $L_N/L$ over time $t$ for different masses for model D in Table 1.}
\end{figure}

\section{Low mass stars that reach the main sequence}

\begin{figure}\label{time needed}
\centering
\includegraphics{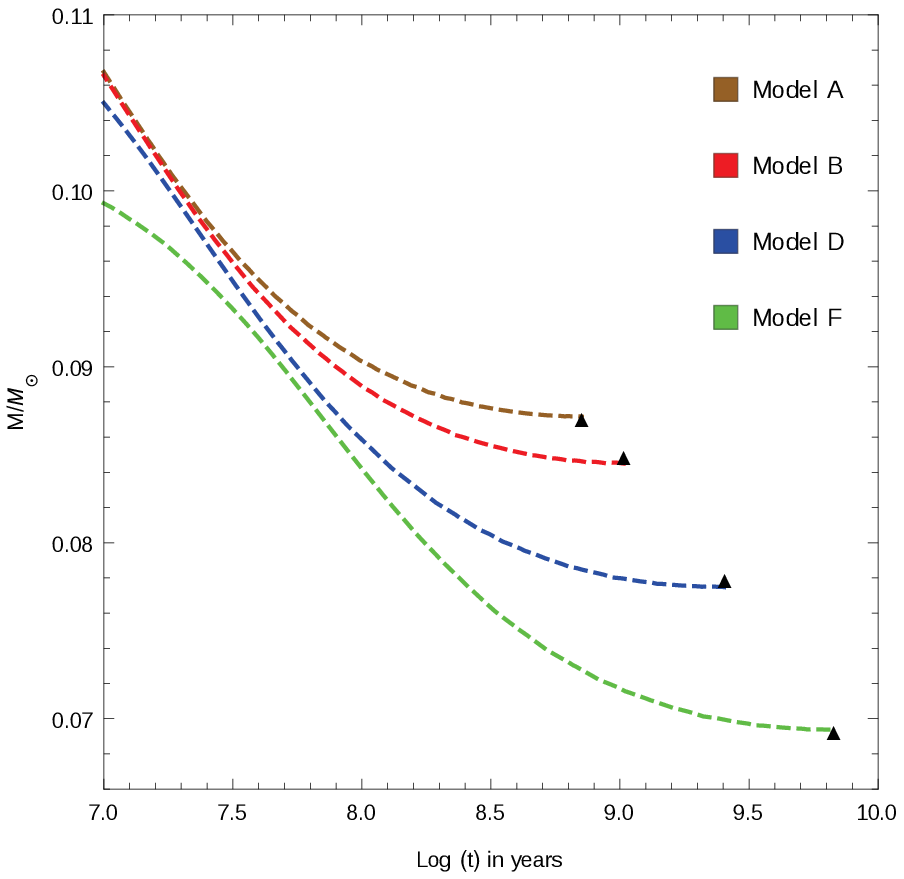}
\figcaption{The time needed to reach the main sequence i.e., $L_N/L =1$, for objects of different masses. Four colored dashed lines represent different models as labeled.The triangles mark the main sequence critical mass for the respective models.}

\end{figure}

In Fig. 5 we plot the time required by low mass stars to reach the main sequence. The curves represent the steady state limit where $L_N/L =1$ for four different models. It is interesting to note that objects of masses at the critical mass boundary between brown dwarfs and main sequence stars, e.g., $0.078 M_{\odot}$ for model D, reach the steady state in about $2.5$ Gyr. This suggests that objects just below the critical mass undergo nuclear burning for an extended period of time but fail to enter the main sequence. Furthermore, stars in the mass range $0.078M_{\odot}-0.086M_{\odot}$ for model D take more than $10^8$ yr to reach the main sequence. Depending on the phase transition points for different models, these numbers vary but the fact that stars close to the minimum mass limit can take an extended amount of time to reach the main sequence means that young stellar clusters may contain a significant fraction of objects that are still in a phase of decreasing luminosity, and behave like brown dwarfs. These objects will ultimately settle into an extremely low luminosity main sequence. 
Here, we estimate the fraction of stars that take more than a specified time to reach the main sequence by using the modified lognormal power-law (MLP) probability distribution function of \citet{Basu14}. Their cumulative distribution function is

\begin{eqnarray}
& F(m)= \frac{1}{2} \rm{erfc}\bigg( -\frac{\rm{ln}(m)-\mu_{0}}{\sqrt{2}\sigma_0}\bigg) \\ \nonumber&-\frac{1}{2} \exp \bigg( \alpha \mu_{o} + \frac{\alpha^2 \sigma_0^2}{2}\bigg) m^{-\alpha} \rm{erfc}\bigg(\frac{\alpha \sigma_0}{\sqrt{2}}- \frac{\rm{ln}(m)- \mu_{0}}{\sqrt{2}\sigma_0}\bigg).
\end{eqnarray}

We use the best fit MLP parameters corresponding to the \citet{chb05} initial mass function (IMF) as obtained by \citet{Basu14}, where $\mu_{0}=-2.404$, $\sigma_{0}=1.044$, and $\alpha =1.396$. We then use the cumulative function to calculate the fraction of stars taking more than either $10^7$ yr, $10^8$ yr or $10^9$ yr to reach the main sequence. Table \ref{tbl:massfraction} contains our results for models A to H. It turns out that about 0.2\% of stars take more than $10^9$ yr to reach the main sequence and about 4\% take longer than $10^8$ yr and about 12\% take longer than $10^7$ yr, for model D. Some of these objects will end up on an extremely low luminosity main sequence, and a sample of luminosity values when an object just above the substellar limit achieves $L_N/L=1$, i.e., reaches the main sequence, is given in Table \ref{tbl:luminosity}.

\begin{deluxetable}{lcccccccc}
\tablecaption{Minimum mass and fractions of stars  \label{tbl:massfraction}} 
\tablehead{
\colhead{Model} & 
\colhead{$M_{\rm min}$\tablenotemark{a} } &
\colhead{$M_9$\tablenotemark{b} } &
\colhead{$N_9$\tablenotemark{c}  } &
\colhead{$M_8$\tablenotemark{b} } &
\colhead{$N_8$\tablenotemark{c} } &
\colhead{$M_7$\tablenotemark{b} } &
\colhead{$N_7$\tablenotemark{c} }  }
\startdata 
A\tablenotemark{d} &0.087 & -         &   -    & 0.090 & 0.014 &  0.107 &  0.085  \\
B &0.085 & 0.085     &  0.000 & 0.089 & 0.020  &  0.107 &  0.095 \\
C &0.081 & 0.081     &  0.001 & 0.087 & 0.029  &  0.106 &  0.108 \\
D &0.078 & 0.078     &  0.002 & 0.086 & 0.038  &  0.105 &  0.118 \\
E &0.073 & 0.075     &  0.006 & 0.085 & 0.052  &  0.103 &  0.127 \\ 
F &0.069 & 0.072     &  0.013 & 0.084 & 0.069  &  0.099 &  0.129 \\
G &0.064 & 0.068     &  0.018 & 0.082 & 0.081  &  0.089 &  0.109 \\
H &0.064 & 0.067     &  0.014 & 0.080 & 0.073  &  0.085 &  0.093 \\
\enddata
\tablenotetext{a}{$M_{\rm min}$ is the minimum mass to reach the main sequence.} 
\tablenotetext{b}{$M_9$, $M_8$ and $M_9$ are the masses up to which the stars take at least $10^9$yr, $10^8$yr and $10^7$ yr, respectively, to reach the main sequence.}
\tablenotetext{c}{$N_9$, $N_8$ and $N_7$ are the number fraction of stars reaching the main sequence in more than $10^9$yr, $10^8$yr and $10^7$ yr, respectively.}
\tablenotetext{d}{Note that for Model A, low mass stars reach main sequence in $< 10^9$ yr. } 
\end{deluxetable}

\begin{deluxetable}{lccccccccccc}
\tabletypesize{\footnotesize} \tablewidth{-4pt} 
\tablecaption{The luminosity at main sequence  \label{tbl:luminosity}} 
\tablehead{
\colhead{$M/M_{\odot}$} & 
\colhead{$ 10^8 (yr)$} &
\colhead{$L/L_{\odot}$ \tablenotemark{a} } }
\startdata 
0.080 & 3.66 & 1.90 $\times 10^{-5}$     \\
0.085 & 1.15 & $9.12 \times 10^{-5}$     \\
0.090 & 0.56 & $2.33 \times 10^{-4}$     \\
0.095 & 0.31 & $4.67 \times 10^{-4}$     \\
\enddata
\tablenotetext{a}{The luminosity of low mass stars in model D when they enter the main sequence.}
\end{deluxetable}


\section{Discussion and conclusions}

This paper presents a simple analytic model of substellar objects. A focus of the paper was to revisit both the development and the shortcomings  of the theoretical understanding of the physics governing the evolution of low mass stars and substellar objects over the last 50 years. We have also made some modifications to the existing models to better explain the physics using analytic forms. Although observational constraints hinder our understanding, a simple analytic model can answer many questions. We have summarized the method of determining the minimum mass for sustained hydrogen burning. Objects in the mass range $0.064M_\odot-0.087M_\odot$ mark this critical boundary between brown dwarfs and the main sequence stars. \\ 
We have derived a general equation of state using  polylogarithm functions \citep{Val10} to obtain the $P-\rho$ relation in the interior of brown dwarfs. The inclusion of the finite temperature correction gives us a much more complete and sophisticated analytic expression of the Fermi pressure (Eq. \ref{pressure eqn3}). The application of this relation can extend to other branches of physics, especially for semiconductor and thermoelectric materials \citep{Molli2011}.\\
The estimate of the surface luminosity is  a challenge given our limitations in understanding the physics inside such low mass objects. Also, it is still an open question if a phase transition actually occurs in a brown dwarf. The results of modern day simulations \citep{Yan15, Mor10} do raise doubts about the relevance of phase transitions in the brown dwarf scenario. We are not aware of well defined analytic models that have a unique way of estimating the surface luminosity apart from using the PPT technique as given in \cite{Burs93}.  In this work, rather than considering a single value for the phase transition point we have used the entire range of temperatures from the phase transition coexistence table \citep{Chab92}. These are within the uncertainty range of the critical temperature of the PPT as proposed by the recent simulations \citep{Yan15}. Thus, considering the large uncertainties involved in such models, this range of values of the minimum mass is much more acceptable than a single distinct transition mass. However, the next step forward is to develop an analytic model for surface temperature that is independent of the PPT.    


We estimate that $\simeq 5 \% $ of stars take more than $10^8$ yr to reach the main sequence, and $\simeq 11 \% $ of stars take more than $10^7$ yr to reach the main sequence (Table 2). The stars in these categories have mass very close to the minimum hydrogen burning limit, and will eventually settle into an extremely low luminosity main sequence with $L/L_{\odot}$ in the range $\approx 10^{-5} - 10^{-4}$. The very low luminosity non-main-sequence hydrogen burning in substellar objects and the pre-main-sequence nuclear burning in very low mass stars are very interesting to study further, and our simplified model can certainly be improved in its ability to estimate the time evolution.




\acknowledgments
We thank Dr. Andrea Becker and Dr. Gilles Chabrier for their valuable comments and input during the preparation of this article.  We also thank Anushrut Sharma for contributing in the initial stages of this project. Our special thanks go to the anonymous reviewer for a thorough critique and constructive suggestions on the manuscript. SB and SRV were supported by a Discovery Grant from the Natural Sciences and Engineering Research Council (NSERC) of Canada. SRV also thanks King's University College for its continued support for his research endeavours.





\newpage

\appendix

\section{Appendix A: Pressure Integral}
The general Fermi Integral can be written as:

\begin{eqnarray}
F_n & = &a \int_{0}^{\infty} \frac{\epsilon^{n}d\epsilon}{e^{\beta(\epsilon-\mu)}+1},\\ \nonumber
& = &a\int_{0}^{\mu} \frac{\epsilon^{n}d\epsilon}{e^{\beta(\epsilon-\mu)}+1}+a\int_{\mu}^{\infty} \frac{\epsilon^{n}d\epsilon}{e^{\beta(\epsilon-\mu)}+1},\\ \nonumber
& = &a\int_{0}^{\mu} \epsilon^{n}d\epsilon-a\int_{0}^{\mu} \epsilon^{n} d\epsilon+a\int_{0}^{\mu} \frac{\epsilon^{n}d\epsilon}{e^{\beta(\epsilon-\mu)}+1}+a\int_{\mu}^{\infty} \frac{\epsilon^{n}d\epsilon}{e^{\beta(\epsilon-\mu)}+1},\\ \nonumber
&= &a\int_{0}^{\mu} \epsilon^{n}d\epsilon-a\int_{0}^{\mu} \frac{\epsilon^{n}d\epsilon}{e^{-\beta(\epsilon-\mu)}+1}+a\int_{\mu}^{\infty} \frac{\epsilon^{n}d\epsilon}{e^{\beta(\epsilon-\mu)}+1}.
\end{eqnarray}

On substituting $ x= -\beta(\epsilon-\mu) $ in the second term and $ x= \beta(\epsilon-\mu) $ in the third term, we arrive at 
\begin{equation}
F_n = a\int_{0}^{\mu} \epsilon^{n}d\epsilon-\frac{a}{\beta}\int^{\beta\mu}_{0}\frac{(\mu -\frac{x}{\beta})^n}{e^x+1}dx + \frac{a}{\beta}\int_{0}^{\infty}\frac{(\mu +\frac{x}{\beta})^n}{e^x+1}dx.
\end{equation}
Let us expand the numerator of the second and the third term of the above integral and retain up to the first three terms: 
\begin{equation}
\left(\mu\pm \frac{x}{\beta}\right)^n\simeq \mu^n\pm n\mu^{n-1}\left(\frac{x}{\beta}\right) + \frac{n(n-1)}{2}\mu^{n-2}\left(\frac{x}{\beta}\right)^2+\ldots .
\end{equation}
Substituting this in the integral for the pressure, we can proceed as follows:
\begin{eqnarray}
F_n =a \int_{0}^\mu \epsilon^{n} d \epsilon+\frac{a}{\beta}\mu^{n}\left\lbrace\int_{0}^\infty\frac{dx}{e^x+1}-\int_{0}^{\beta\mu}\frac{dx}{e^x+1}\right\rbrace \\ \nonumber
+n\frac{a}{\beta^2}\mu^{n-1}\left\lbrace\int_{0}^\infty\frac{xdx}{e^x+1}+\int_{0}^{\beta\mu}\frac{xdx}{e^x+1}\right\rbrace\\ \nonumber
-\frac{n(n-1)}{2}
\frac{a}{\beta^3}\mu^{n-2}\left\lbrace\int_{0}^{\beta\mu}\frac{x^2dx}{e^x+1}-\int_{0}^\infty\frac{x^2dx}{e^x+1}\right\rbrace.
\end{eqnarray}

Substituting $n=\frac{3}{2}$ for the Fermi pressure (Eq. \ref{eqn:pressure equation2}) and evaluating these integrals using the polylogs \citep{Val10} we arrive at a simplified form 
\begin{equation}
P_F\simeq a\frac{2}{5}\mu^{\frac{5}{2}}-\frac{1}{8}a\beta^{-1}\mu^\frac{3}{2}\ln(1+e^{-\beta\mu})+\frac{3}{2} \frac{\pi^2}{6}a\beta^{-2}\mu^\frac{1}{2}+\frac{3}{4}a\beta^{-2}\mu^\frac{1}{2}\Li_2(-e^{-\beta\mu})-\frac{3}{4}a\beta^{-3}\mu^{-\frac{1}{2}}\Li_3(-e^{-\beta\mu})\ldots
\end{equation}
Similarly for $n=\frac{1}{2}$, the expression for the number density can be obtained as
\begin{equation}\label{rhogen}
\rho\simeq a\frac{2}{3}\mu^{\frac{3}{2}}+\frac{3}{8}a\beta^{-1}\mu^{\frac{1}{2}}\ln(1+e^{-\beta\mu})+\frac{\pi^2}{12}a\beta^{-2}\mu^{-\frac{1}{2}}  +\frac{3}{4}a\beta^{-2}\mu^{-\frac{1}{2}} \Li_2(-e^{-\beta\mu})+\frac{1}{4}a\beta^{-3}\mu^{-\frac{1}{2}}\Li_3(-e^{-\beta\mu})\ldots
\end{equation}

\section{Appendix B: Surface Properties}
The surface properties of a brown dwarf can be analyzed by studying the phase diagram of hydrogen in the interior and the photospheric region. The total pressure inside a stellar or substellar object can be represented as
\begin{equation}
P= P_g+P_r,
\end{equation}
where $P_g$ is the gas pressure due the the adiabatic ideal gas and $P_r$ is the radiation pressure. At a temperature comparable to that of the envelope surrounding the interior of a substellar object, the hydrogen is partially ionized and the helium gas is mostly molecular. For a quasistatic change, the first law of thermodynamics yields 
\begin{equation}\label{1st law}
dQ = \left(\frac{\partial U}{\partial T}\right)_v dT+\left(\frac{\partial U}{\partial V}\right)_TdV + P dV
\end{equation}
\citep{Clay68}. The most general expression for the internal energy of a monatomic gas is
\begin{equation}\label{the total energy}
U= \frac{3}{2}\frac{k_B N_0 T}{ \mu_1}+aT^4V+ x \chi N_0,
\end{equation}
where we have considered the energy due to photon radiation and gas ionization. Here  $\chi$ is the ionization energy and $x$ is the ionization fraction of hydrogen. Taking the partial derivatives of Eq. (\ref{the total energy}) and using the second law of thermodynamics $TdS = dQ$ we arrive at
\begin{equation}\label{The entropy}
dS = -\frac{3}{2} \frac{k_B N_0}{\mu_1}\frac{dT}{T}+ \frac{k_B N_0}{\mu_1} \frac{dW}{W}+\frac{4}{3}a\, d\,W +N_0 \left( \chi + \frac{3}{2}k_BT\right)\left(\frac{\partial x}{\partial T}\frac{dT}{T}+\frac{\partial x}{\partial V}\frac{dV}{T} \right) .
\end{equation} 
where $W=T^3 V $.
The ionization fraction is a function of density and temperature, $x(\rho,T)$. Using the Saha equation we obtain 
\begin{equation}\label{saha1}
\left(\frac{\partial x}{\partial T}\right)_V=\frac{x(1-x)}{2-x}\frac{1}{T}\left(\frac{3}{2}+\frac{\chi}{k_BT} \right),
\end{equation} 
\begin{equation}\label{saha2}
\left(\frac{\partial x}{\partial V}\right)_T=\frac{x(1-x)}{2-x}\frac{1}{V}\,.
\end{equation} 
Using the above relations, we can simplify Eq. (\ref{The entropy}) to be
\begin{equation}\label{general entropy appn}
\frac{dS}{k_B N_0} = -\frac{3}{2\mu_1}  \frac{dT}{T}+  \frac{dW}{\mu_1 W} + \left(\frac{3}{2}\right)^2 \frac{HdT}{T}+ \frac{3}{2} \frac{HdV}{V}+\frac{4adW}{3k_B N_0}+ \chi \left(\frac{dV}{TV}+3\frac{dT}{T^2}+\frac{\chi}{k_B}\frac{dT}{T^3}\right),
\end{equation}
where $H=\frac{x(1-x)}{2-x}$. This expression is the same as Eq. (\ref{dq}) except for the final two terms due to radiation and ionization energy, respectively. We can ignore the final term and retain terms of linear order in $T$.   
On integrating and simplifying the above expression we get the entropy for the partially ionized hydrogen and helium gas, 
\begin{equation}
S_1 = k_B N_0\left(\frac{1}{\mu_1}+\frac{3}{2}\frac{x(1-x)}{2-x}\right) \ln \frac{T^\frac{3}{2}}{\rho} +\frac{4}{3}a T^3V +  C_1.
\end{equation}

At low temperature ($T<4000K$) and low pressure, hydrogen is predominantly molecular and fluid. Repeating the above derivation for the non-ionized diatomic hydrogen gas with energy $U=\frac{5}{2}\frac{k_BTN_0}{\mu_2}$ and molecular helium we can arrive at an expression for the entropy, 
\begin{equation}\label{diatomic entropy}
S_2 = \frac{k_B N_0}{\mu_2} \ln \frac{T^\frac{5}{2}}{\rho} +\frac{4}{3}a T^3V + C_2.
\end{equation}

In the above expressions for entropy, $\rho$ and $T$ are the density and the temperature, respectively. Other variables are as described in the text.

\section{Appendix C: Thermal Properties}
We discuss some of the more accurate expressions for the important thermal properties of a degenerate system.
\subsection{Fermi energy}
Using Eq. (\ref{rhogen}) for the number density  we can write the general expression for the chemical potential $\mu$ in terms of the Fermi energy $\mu_F$ at $T=0$ \citep{Feyn72}.
Considering only the first three terms Eq. (\ref{rhogen}) and for $\rho =\frac{2}{3}\mu_F^{\frac{3}{2}}$, we find 
\begin{equation}
\mu\simeq \mu_F -\frac{\pi^2}{12}\frac{1}{(\beta\mu_F)^2}-\frac{3}{8}\frac{1}{\beta\mu_F}\ln(1+\exp^{-\beta\mu_F}).
\end{equation}
The second and the third terms are the correction factor $C$ to the zero temperature Fermi energy.
For $0.03 <  \frac{1}{\mu_F \beta}<0.20$ the correction factor $C$ will be in the range $\sim 8\times 10^{-4}<C<4 \times 10^{-2}$.\\
\subsection{Specific Heat}
In the nondegenerate completely ionized limit, the specific heat $C_v\sim 3k_BN/2$. At finite temperatures, the value of the specific heat is less than the limiting value, i.e., $C_v< \frac{3Nk_B}{2}$. The specific heat of the ideal Fermi gas decreases monotonically. At low but finite temperatures,
\begin{equation}
\frac{C_v}{N}\simeq \frac{\pi^2}{2}\frac{k_BT}{\mu_F}.
\end{equation}
A detailed analysis in the calculation of the specific heat shows that the numerical coefficients in the expansion approached a limiting value of 2.
\subsection{Grueneisen Parameter}
Applying the condition of constant entropy to Eq. (\ref{the interior entropy}) leads to the condition
\begin{equation}\label{Grcnd}
T=C\rho^{\frac{2}{3}},
\end{equation}
where $C$ is a constant. The Grueneisen parameter $\gamma$ is given by the expression 
\begin{equation}
\gamma = \left(\frac{\partial\,log \,T}{\partial\, log\,\rho}\right) _s .
\end{equation}
Using Eq. (\ref{Grcnd}) in the above expression we estimate the value of the Grueneisen parameter $\gamma $ to be $\frac{2}{3}$. This value is in approximate accord with \cite{Ste91}, who indicated that $\gamma \simeq0.6$ in a dense Coulomb plasma when obtained from computer simulations.

\subsection{Ionic correlation}
Ionic correlation is an important contribution as considered by \cite{St096}, \cite{Beck15}, \cite{Hubb84} and \cite{Ger10} to name a few. \cite{St096} use the method of Pade's approximations to provide  explicit expressions for the fully ionized plasma of the Helmholtz free energy and the pressure. They have considered the nonideal effects of different correlations such as the electron-electron, ion-ion, electron-ion, as well as exchange contribution for a wide range of values of the Coulomb coupling parameter $\Gamma$, which is the ratio of the Coulomb to thermal energy:
\begin{equation}
\Gamma = \frac{e^2}{k_B T} \left(\frac{4\pi n_e}{3}\right)^{\frac{1}{3}} = \frac{\Gamma_{\rm ion} }{\langle z^{\frac{5}{3}}\rangle}.
\end{equation}
Here $n_e$ stands for the electron density and $\langle z^{\frac{5}{3}}\rangle$ is the charge average, given as
\begin{equation}
\langle z^{\frac{5}{3}}\rangle= \frac{\sum n_i Z_i^{\frac{5}{3}}}{\sum n_i},
\end{equation}
for ions of different species $i$.
The greatest effect is in the relative pressure contribution $P_{ii}/P_{\rm ideal}$ of the ionic correlation term for hydrogen at $T= 10^5 \, \rm K$,  estimated to be  $\sim -\,0.1$
at $\rho \sim 10^3\, \rm g/cm^3$ and a minimum of $\sim -\,0.2$ at $\rho \sim 10 \, \rm g/cm^3$.



\clearpage




\end{document}